\documentclass[aps,11pt]{revtex4}
\usepackage{epsfig}
\usepackage{amsmath}
\usepackage{bm}
\usepackage{times}

\def\beq{\begin{equation}}
\def\eeq{\end{equation}}
\def\bea{\begin{eqnarray}}
\def\eea{\end{eqnarray}}
\def\nn{\nonumber}
\def\Eq#1{Eq.~(\ref{#1})}
\def\gut{\mathrm{GUT}}

\begin{document}
\preprint{IFIC/06-33} \vspace{5.0cm}
\title{ON UNIFICATION AND NUCLEON DECAY IN SUPERSYMMETRIC
GRAND UNIFIED THEORIES BASED ON $SU(5)$}
\author{Ilja Dor\v{s}ner$^{1}$}
\email{idorsner@phys.psu.edu}
\author{Pavel Fileviez P\'erez$^{2}$}
\email{fileviez@cftp.ist.utl.pt}
\author{Germ\'an Rodrigo$^{3}$}
\email{german.rodrigo@ific.uv.es}
\affiliation{
$^{1}$The Pennsylvania State University \\
104 Davey Lab, PMB 025, University Park, PA 16802
\\
$^{2}$Centro de F{\'\i}sica Te\'orica de Part{\'\i}culas \\
Departamento de F{\'\i}sica.\ Instituto Superior T\'ecnico \\
Avenida Rovisco Pais 1, 1049-001 Lisboa, Portugal
\\
$^{3}$Instituto de F\'{\i}sica Corpuscular,
CSIC-Universitat de Val\`encia \\
Apartado de Correos 22085, E-46071 Valencia, Spain}
\date{\today}
\begin{abstract}
We investigate the unification constraints in the minimal sypersymmetric 
grand unified theories based on $SU(5)$ gauge symmetry. The most general 
constraints on the spectrum of minimal supersymmetric $SU(5)$ and 
flipped $SU(5)$ are shown. The upper bound on the mass of the colored 
Higgs mediating proton decay is discussed in detail in the context of 
the minimal supersymmetric $SU(5)$. In the case of the minimal 
SUSY $SU(5)$ we show that if we stick to the strongest bound on 
the colored triplet mass coming from dimension five proton 
decay contributions there is no hope to test this model 
at future nucleon decay experiments through the dimension six operators. 
We find a lower bound on the partial proton decay lifetime for all 
relevant channels in the context of flipped SUSY $SU(5)$. We conclude 
that flipped SUSY $SU(5)$ might be in trouble if proton decay is 
found at the next generation of experiments with a lifetime 
below $10^{37}$ years. 
\end{abstract}
\maketitle
\pagestyle{plain}
\section{Introduction}
The smallest special unitary group that allows embedding of the Standard Model (SM) 
is $SU(5)$~\cite{GG}. On account of its unique symmetry breaking pattern $SU(5)$ 
represents a fundamental framework to test the idea of grand unification. 
It is thus necessary to have clear understanding of its generic predictions 
and testable consequences.

The most dramatic prediction of grand unified theories by far is the
decay of the proton~\cite{Pati-Salam,Langacker,review}. This generic 
feature has accordingly been a prime target of numerous experimental searches
but a positive signal has not yet been observed. Nevertheless, existing
experimental limits on the proton decay lifetime have already placed
rather severe bounds on simple models of grand unification. This is
especially applicable to the minimal models based on $SU(5)$ which
will be our preferred framework of interest. We will in particular
focus on the current status of these models---in their supersymmetric
form---with respect to the latest experimental results on proton decay.

As is well-known non-supersymmetric grand unified theories, in
particular the simplest realizations of the ordinary $SU(5)$, 
were considered for a long time to be ruled out by proton decay.
However, it was shown recently that there still exists viable
parameter space in minimal realistic models that is yet to 
be excluded~\cite{non-SUSY-SU(5),non-SUSY-SU(5)I}. Also, 
it has turned out that proton decay might be absent altogether 
in a class of models based on flipped $SU(5)$~\cite{rotating}. 
Both results draw from a recent study that focused on all 
relevant contributions to proton decay in non-supersymmetric 
GUTs~\cite{upper}.

The situation with the status of the minimal supersymmetric
$SU(5)$~\cite{SUSYSU(5)} grand unified theory is more involved. This
is primarily due to the fact that supersymmetry generates a large
number of model dependent proton decay contributions in addition
to the non-supersymmetric ones making general analysis intricate.
In the context of flipped SUSY $SU(5)$ the dimension six gauge 
contributions for proton decay are the most important. Therefore, 
there is no problem to satisfy the current experimental bound on 
the proton lifetime.    

In this work we study the unification constraints in the context of
minimal supersymmetric models based on $SU(5)$ and the corresponding 
implications for proton decay. To be as broad as possible we will 
look at both types of minimal supersymmetric matter unification 
under $SU(5)$---the ordinary and flipped one.
In minimal supersymmetric $SU(5)$ we investigate the case when the
fields in the adjoint representation $\Sigma_3$
(triplet of $SU(2)$) and $\Sigma_8$ (octet of $SU(3)$) are not 
degenerated. In flipped SUSY SU(5) we study the most general unification 
constraints. The constraints coming from proton decay are discussed 
in both models. We find a lower bound on the proton decay lifetime 
in flipped SUSY $SU(5)$ and conclude that it will be very difficult 
to test both models, minimal SUSY $SU(5)$ and flipped SUSY $SU(5)$, 
at future proton decay experiments through the dimension six operators. 

Our work is organized as follows: in Section~1 we show the most 
general unification constraints in minimal supersymmetric $SU(5)$ and 
their implications for proton decay predictions. In Section~2 we 
present the relevant predictions in the framework of the minimal 
supersymmetric version of flipped $SU(5)$. Finally, we summarize 
our results in Section III.

\section{Minimal SUSY $SU(5)$: Unification versus nucleon decay}
Minimal SUSY $SU(5)$~\cite{SUSYSU(5)} unifies one
generation of matter of the Minimal Supersymmetric Standard Model
(MSSM) in two superfields $\mathbf{\hat{\bar 5}} =
(\hat{d}^C,\hat{L})$ and $\mathbf{\hat{10}} =
(\hat{u}^C,\hat{Q},\hat{e}^C)$, while its Higgs sector comprises
$\mathbf{\hat 5_H}=(\hat{T}, \hat{H})$, $\mathbf{\hat{\bar
5}_H}=(\hat{\bar T},\hat{\bar H})$, and $\mathbf{\hat{24}_H}$. In
our notation the SM decomposition of the adjoint Higgs superfield
reads $\mathbf{\hat{24}}=(\hat{\Sigma}_8, \hat{\Sigma}_3,
\hat{\Sigma}_{(3,2)}, \hat{\Sigma}_{(\bar{3}, 2)},
\hat{\Sigma}_{24})=(\bm{8},\bm{1},0)+(\bm{1},\bm{3},0)+(\bm{3},\bm{2},-5/6)
+(\overline{\bm{3}},\bm{2},5/6)+(\bm{1},\bm{1},0)$. 
In addition, it needs to accommodate at least two $SU(5)$ singlet 
superfields in order to generate observed neutrino masses or 
use the bilinear R-parity violating interactions. 

It has been claimed that this theory, in its renormalizable form
and with \textit{low-energy SUSY}, was excluded on the proton
decay grounds~\cite{Murayama}. For previous studies see
~\cite{Nath-pd1, Nath-pd2, Hisano}.
More precisely, it has been shown~\cite{Murayama} that 
in order to satisfy the experimentally established lower bound on 
proton lifetime the mass $M_T$ of the triplet fields $\hat{T}$ 
and $\hat{\bar T}$ had to be greater than the upper bound 
on $M_T$ that was extracted from requirement to have successful 
gauge coupling unification. However, it is
well-known that the minimal renormalizable SUSY $SU(5)$ is not
realistic since it is not possible to establish phenomenologically
consistent fermion masses and mixings within its framework. So,
the above claim seems redundant to say the least. In fact, any
study that aims to rule out any given GUT on the proton decay
grounds should be undertaken within a realistic scenario for
fermion masses and mixings. We will adhere to this principle in
our study.

In order to keep SUSY $SU(5)$ minimal but realistic with respect
to the quark and lepton mass spectrum it is sufficient to take
into account nonrenormalizable operators in the Yukawa
sector~\cite{Ellis:1979fg}. These operators modify the bad
relation $Y_D=Y_E^T$ in a way that renders theory realistic. As
usual, $Y_D$ ($Y_E$) is the down quark (charged lepton) Yukawa
matrix. Once these operators are present the couplings of the
triplet Higgs to matter are also modified. In fact, one can even 
set to zero all the couplings of the triplets to matter~\cite{Dvali}. 
This in turn practically removes any phenomenological bound on 
the triplet mass. It is thus clear that it is rather difficult 
to rule out the entire parameter space of the minimal SUSY $SU(5)$. 
This issue has been studied in detail in reference~\cite{Bajc1,Bajc2}, 
while the impact of the higher-dimensional operators on proton decay 
has been studied in~\cite{Nath1,Nath2,Berezhiani,Bajc2,David}.

If the higher-dimensional operators are allowed in the Yukawa
sector one should for consistency also consider other possible
nonrenormalizable operator contributions and investigate their
impact on the viability of the theory. We refer to two additional
types of operators in particular. The first type modifies the mass
spectrum of the Higgs fields responsible for the GUT symmetry
breaking with respect to the renormalizable case~\cite{Bajc2}. The
second type affects boundary conditions for the gauge coupling
unification through modification of the gauge kinetic
terms~\cite{Shafi}. It is our intention to investigate in detail
influence of the first type of operators on the predictions of the
theory. As we will show, these modifications alone are sufficient
to make theory realistic with respect to the proton decay
constraints. This relaxes the need to fine-tune relevant Yukawa
couplings to simultaneously recreate observed masses and mixings
and suppress couplings of the triplet to matter. Preliminary study
of their impact has already been presented in Ref.~\cite{Bajc2}.
Our analysis will not only be more detailed but will also reflect
recent improvements in our knowledge of low-energy data as given
in Ref.~\cite{pdg}. In addition, we will investigate consequences
of particular realizations of the SUSY spectrum on the proton
decay predictions. It should finally be mentioned that even the
second type of operators is self sufficient in rendering theory
realistic with respect to proton decay bounds~\cite{review}.

\subsection{Unification constraints: octet-triplet splitting}
The mass splitting between octet ($\Sigma_8$) and triplet
($\Sigma_3$) of adjoint Higgs superfield as the simplest way to
satisfy conservative experimental bound on $M_T$ within the
minimal SUSY $SU(5)$ framework has first been suggested in
Ref.~\cite{Bajc2}. The idea is based on the fact that $M_T$ scales
as $(M_{\Sigma_3}/M_{\Sigma_8})^{5/2}$ if successful gauge
coupling unification at the one-loop level is imposed. Hence,
sufficiently strong $\Sigma_8$-$\Sigma_3$ mass splitting could
lead to $M_T$ being heavy enough to avoid even the most
conservative constraints coming from experimental results on
partial proton decay lifetimes. However, this approach cannot be
implemented in renormalizable theory where $M_{\Sigma_8}=M_{\Sigma_3}$. 
To break this degeneracy one needs to consider presence of 
nonrenormalizable contributions which are anyhow necessary in 
order to accommodate observed masses of quarks and 
leptons~\cite{Ellis:1979fg}.

To outline how this idea works let us consider the superpotential up
to the first order in $\langle \Sigma \rangle/\Lambda$. It reads
\begin{equation}
\label{superpotential} W_\Sigma =  m \, \rm{Tr} \hat{\Sigma}^2 +
\lambda \, \rm{Tr} \hat{\Sigma}^3 +  \frac{a}{\Lambda} \, \left(
\rm{Tr} \hat{\Sigma}^2\right)^2 +  \frac{b}{\Lambda} \, \rm{Tr}
\hat{\Sigma}^4~,
\end{equation}
where $\langle \Sigma \rangle$ is a vacuum expectation value (VEV)
of the order of the GUT scale ($M_{\gut}$) while $\Lambda$ can be
identified with the scale of gravity ($M_\mathrm{Planck}$). If
only the first two terms are taken into consideration one obtains
$M_{\Sigma_3}=M_{\Sigma_8}$ if $\langle \Sigma \rangle$ points in
the SM direction. This hinders the possibility of increasing $M_T$
to arbitrarily high scale since, as we said before, $M_T$ depends
on $M_{\Sigma_3}$ and $M_{\Sigma_8}$ only through their ratio. If,
on the other hand, one considers a more general scenario---when
all the terms in Eq.~\eqref{superpotential} are taken into
account---it is possible to have very wide range of values for the
ratio in question. For example, if one neglects for simplicity the
term proportional to $\lambda$ in Eq.~\eqref{superpotential} one
obtains $M_{\Sigma_3}=4 M_{\Sigma_8}$~\cite{Bajc2}. That is more
than sufficient to bring even the most conservative predictions of
the minimal SUSY $SU(5)$ in agreement with experimental findings
as we show later.

To explicitly show how $\Sigma_3$-$\Sigma_8$ mass splitting enters
into prediction for $M_T$ we must resort to renormalization group
equations for the SM gauge couplings. At the one-loop level, they
are given by:
\bea 
\alpha_1^{-1} (M_Z) &=& \alpha_\gut^{-1} + \frac{1}{2\pi}
\left(
   \frac{41}{10} \ln \frac{M_\gut}{M_Z}
+  \frac{21}{10} \ln \frac{M_\gut}{M_{\tilde q}} +  \frac{2}{5}
\ln \frac{M_\gut}{M_{\tilde G}} -  10            \ln
\frac{M_\gut}{M_V}
+ \frac{2}{5}    \ln{\frac{M_\gut}{M_T}} \right), \nn 
\eea
\bea
\alpha_2^{-1} (M_Z) &=& \alpha_\gut^{-1} + \frac{1}{2\pi} \left( -
\frac{19}{6}  \ln \frac{M_\gut}{M_Z} +  \frac{13}{6}  \ln
\frac{M_\gut}{M_{\tilde q}} +  2             \ln
\frac{M_\gut}{M_{\tilde G}} -  6             \ln
\frac{M_\gut}{M_V}
+  2             \ln \frac{M_\gut}{M_{\Sigma_3}} \right), \nn 
\eea
\bea
\alpha_3^{-1} (M_Z) &=& \alpha_\gut^{-1} + \nn \\ &+&
\frac{1}{2\pi} \left( - 7             \ln \frac{M_\gut}{M_Z} +  2
\ln \frac{M_\gut}{M_{\tilde q}} +  2             \ln
\frac{M_\gut}{M_{\tilde G}} -  4             \ln
\frac{M_\gut}{M_V} +                \ln{\frac{M_\gut}{M_T}}
+  3             \ln \frac{M_\gut}{M_{\Sigma_8}} \right). \nn \\
\label{RGEsusy} 
\eea
Here, for simplicity we assume the same mass $M_{\tilde q}$ 
for all MSSM scalars, i.e., sfermions and the extra Higgs doublet, 
and the same mass $M_{\tilde G}$ for Higgsinos and gauginos. 
We comment on a more general scenario that accommodates the 
splitting between the relevant gaugino masses later on. 
As usual, $M_V$ is the mass of superheavy gauge bosons 
while $M_{GUT}$ represents the scale where gauge couplings 
unify. We assume $M_V=M_{GUT}$ in what follows which is 
especially justified in the two-loop analysis which we also 
present towards the end of this section.

It is easy to solve for $M_T$ in terms of all other mass scales
that appear in~\Eq{RGEsusy}. If we eliminate $\alpha_\gut$ we end
up with two equations. They read
\bea  
\label{tripletmass} 
M_T &=&
\left(\frac{M_{\Sigma_3}}{M_{\Sigma_8}} \right)^{5/2}
(M_{\tilde{G}}^4 M_{\tilde{q}} M_Z)^{1/6} \exp \left[-\frac{5}{6}
\pi \left( \alpha_1^{-1} (M_Z)  -  3 \alpha_2^{-1} (M_Z)  +  2
\alpha_3^{-1} (M_Z) \right)
\right]~, \\
\frac{M_{\Sigma_3}}{M_{\Sigma_8}} &=&
\frac{M_Z^{22/3}}{M^2_{\Sigma_8} M_{\tilde{G}}^{4/3} M_V^4} \exp
\left[\frac{\pi}{3} \left( 5\alpha_1^{-1}(M_Z)  -  3
\alpha_2^{-1}(M_Z)  -  2 \alpha_3^{-1}(M_Z) \right) \right]~.
\label{m3} 
\eea 
Of course, $M_T$ and $M_V$ are then related through 
\beq 
\label{second} 
M_T = \frac{M_{\tilde{q}}^{1/6} \, M_Z^{37/2}} 
{M_{\Sigma_8}^5 \, M_{\tilde{G}}^{8/3}  \, M_V^{10}}
\, \exp \left[\frac{10\pi}{3} \left( \alpha_1^{-1}(M_Z)-
\alpha_3^{-1}(M_Z) \right) \right]~. 
\eeq
Clearly, $M_T$ scales as $(M_{\Sigma_3}/M_{\Sigma_8})^{5/2}$~\cite{Bajc2}. 
Thus, the larger the $M_{\Sigma_3}/M_{\Sigma_8}$ ratio is the 
larger $M_T$ becomes as we initially suggested. If this ratio is 
set to one then Eq.~\eqref{tripletmass} simplifies and $M_T$ can 
consequently be easily constrained by low energy input and 
assumptions with regard to the SUSY spectrum~\cite{Barbieri}.
This in turn implies that the triplet mass is too light to satisfy experimental 
constraints~\cite{Murayama} from proton decay if one neglects the higher 
dimensional operators and the quark and lepton mixings. In this case we get 
the strongest bound on the colored triplet mass. So, how much should 
the ratio depart from one if we want $M_T$ to be above the most 
conservative experimental bound?

To answer that we first update the result of Ref.~\cite{Murayama}
according to which the current bound on the partial proton
lifetime---$\tau(p \to K^+ \bar{\nu})
> 2.3 \times 10^{33}$\,years---implies the following bound on the
triplet mass: $M_T > 1.4 \times 10^{17}$\,GeV. Using this
conservative constraint, $M_{\tilde{G}}=M_{\tilde{q}}=M_Z$, and
$M_V=M_\gut$ we obtain from~\Eq{tripletmass} the one-loop result
\beq \label{borut} M_{\Sigma_3} > 2.0 M_{\Sigma_8}. \eeq So, if
one allows for the presence of nonrenormalizable contributions in
the superpotential one can certainly make minimal SUSY
$SU(5)$~\cite{Bajc2} realistic as long as Eq.~\eqref{borut} holds
without the need to suppress couplings of the triplet to the
matter. In fact, since there are two relevant equations, we obtain
an additional constrain. Namely, Eq.~\eqref{second} simultaneously
implies $M_{\Sigma_8} < 3.0 \times 10^{13}$\,GeV in order that
$M_{GUT} \geq M_T$.

\begin{figure*}
\begin{center}
\epsfig{file=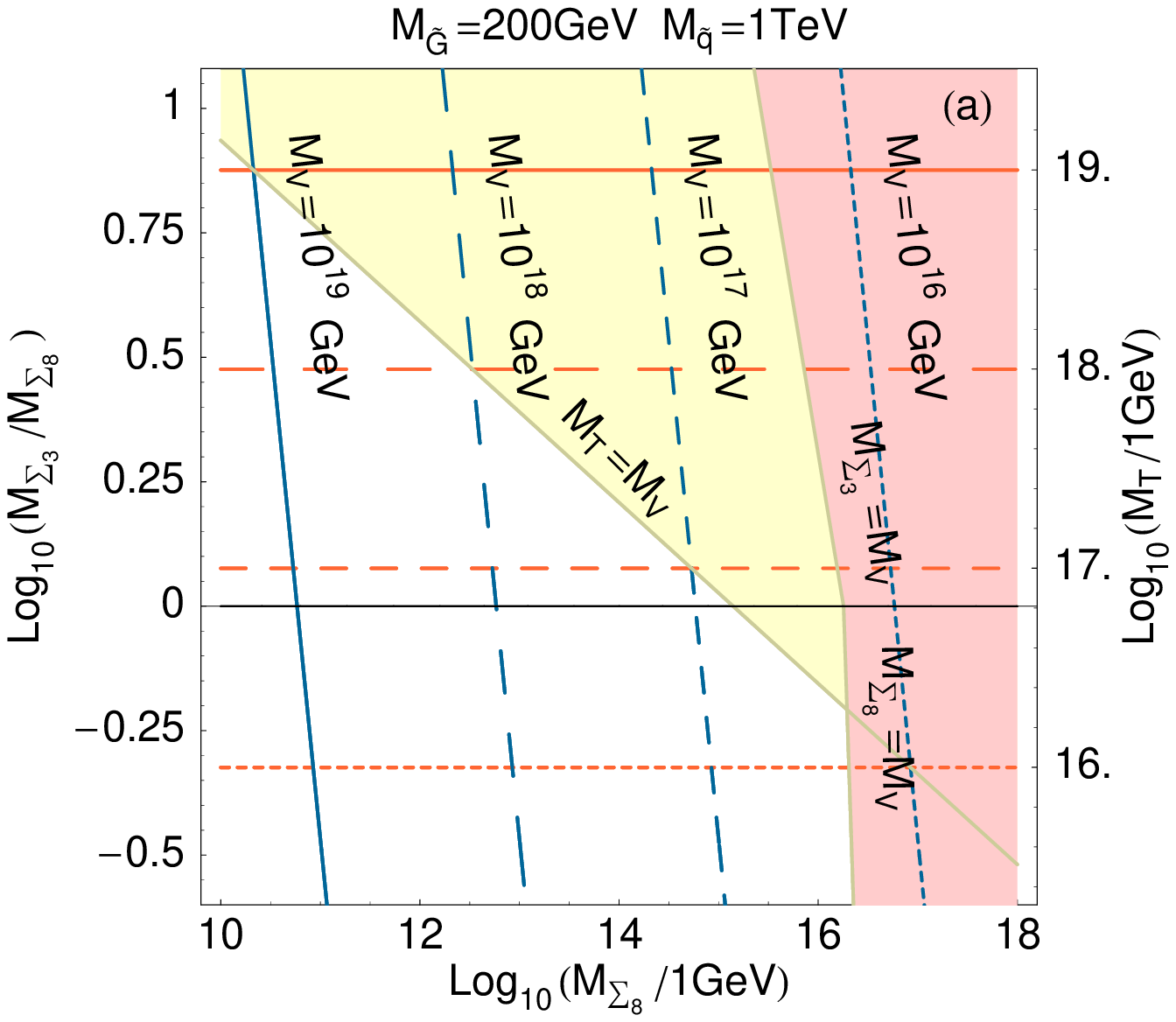,width=8cm}
\epsfig{file=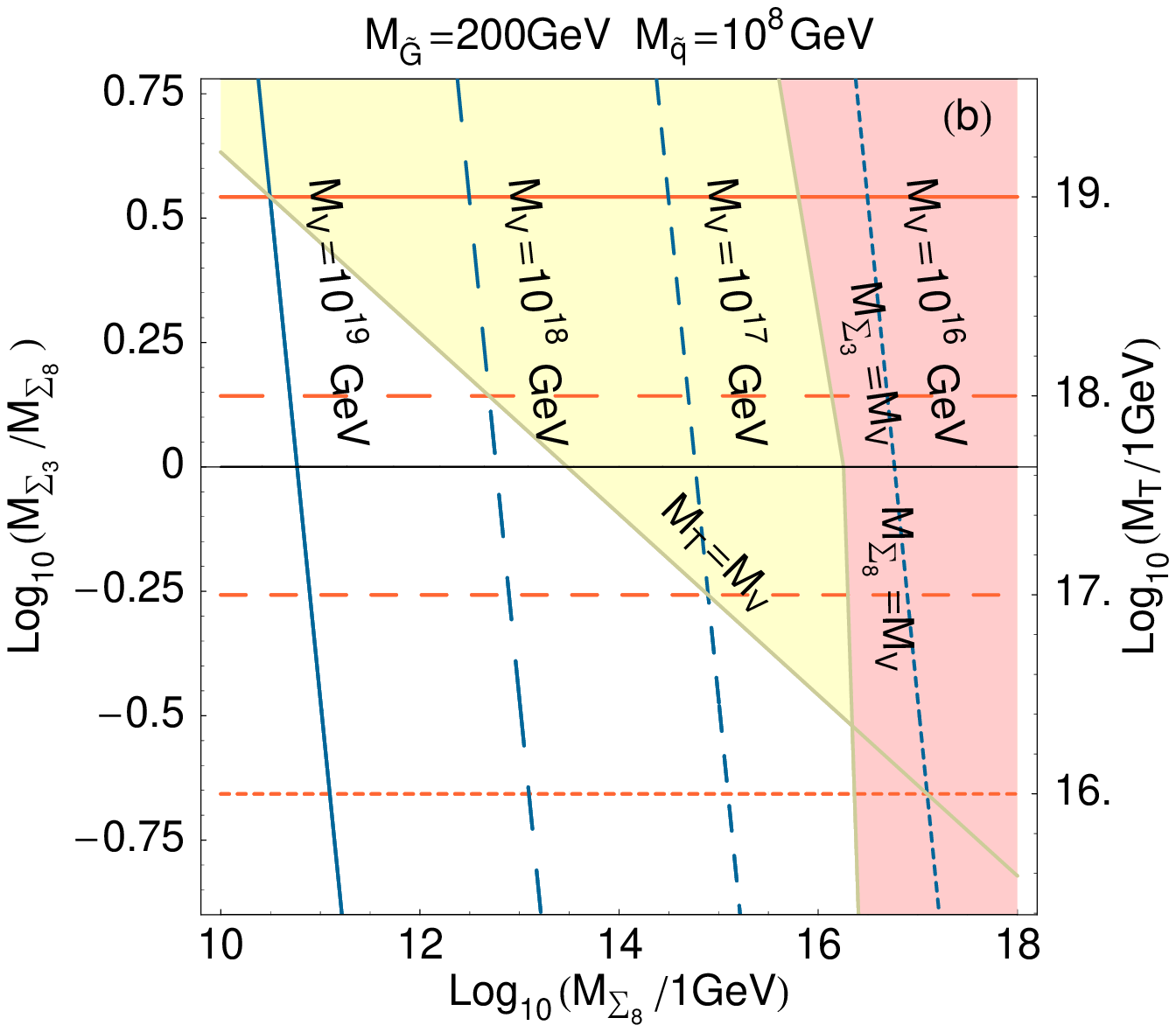,width=8cm}
\epsfig{file=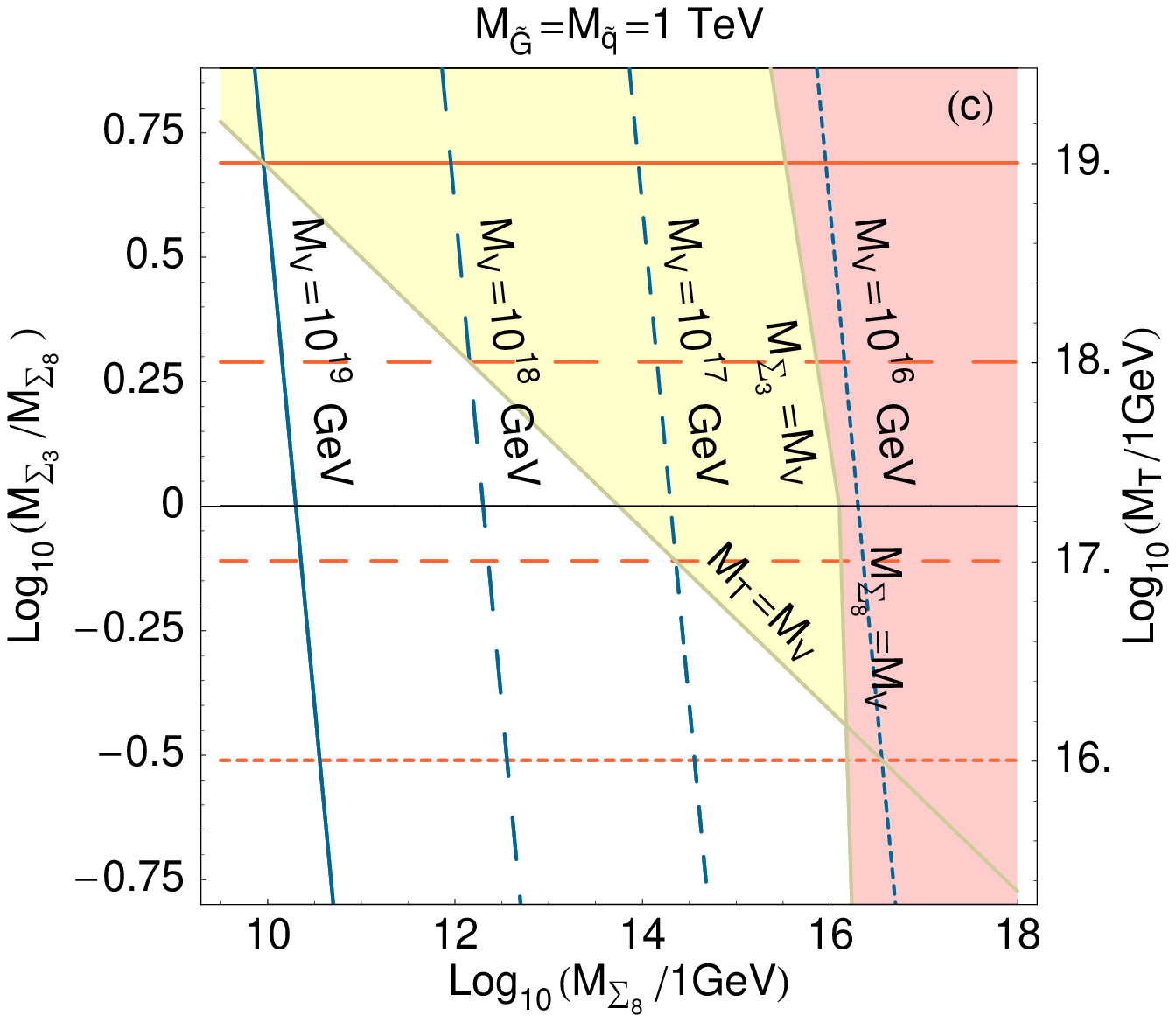,width=8cm} 
\caption{Parameter space for the gauge coupling unification in the
$M_{\Sigma_3}$-$M_{\Sigma_8}$ plane for different values of
$M_{\tilde q}$ and $M_{\tilde G}$ in the $\overline{DR}$ scheme.
Lines of constant $M_T$ and $M_V=M_{GUT}$ are shown. The light
shaded area is excluded by the constraint $M_T \leq M_V$ while the
dark shaded area is excluded by either $M_{\Sigma_3} \leq
M_V=M_{GUT}$ or $M_{\Sigma_8} \leq M_V=M_{GUT}$. As input
parameters we take $\alpha_s (M_Z)_{\overline{MS}}=0.1176$,
$\sin^2 \theta_W (M_Z)_{\overline{MS}}=0.2312$ and
$\alpha(M_Z)_{\overline{MS}}= 1/127.906$.} \label{fig:parameter1}
\end{center}
\end{figure*}

Let us now study unification constraints on the spectrum of the
minimal realistic SUSY $SU(5)$ in detail. Using~\Eq{tripletmass}
and \Eq{m3} we can plot the parameter space allowed by unification
in the $M_{\Sigma_3}$-$M_{\Sigma_8}$ plane for fixed values of
$M_T$ and $M_V=M_{GUT}$. The whole parameter space is shown in
Fig.~\ref{fig:parameter1} assuming different values of $M_{\tilde
G}$ and $M_{\tilde q}$. The allowed region in the context of the
minimal SUSY $SU(5)$ as shown in Fig.~\ref{fig:parameter1} is the
region bounded from above by $M_T \le M_V=M_{\gut}$, from the left
by $M_V=M_{\gut}<M_\mathrm{Planck}$ and from the right by
$M_{\Sigma_8}\le M_V=M_{\gut}$. The constraint $M_{\Sigma_3}\le
M_V=M_{\gut}$ does not play any role due to the other exclusion
limits. In Fig.~\ref{fig:parameter1}a we show the possibility to
achieve unification for the case $M_{\tilde G}=200$\,GeV and
$M_{\tilde q}= 1$\,TeV, while in Fig.~\ref{fig:parameter1}b the
corresponding parameter space for the so-called ``Split SUSY''
scenario~\cite{splitsusy} is shown. To implement relevant
experimental bounds on the masses of SUSY particles we use
Ref.~\cite{pdg}. Clearly, the allowed region for
$M_{\Sigma_3} > M_{\Sigma_8}$ in Fig.~\ref{fig:parameter1}b has been
considerably reduced with respect to Fig.~\ref{fig:parameter1}a.
In all those plots we observe the well known fact that the
contributions of the fermionic superpartners are very important
for unification, while the contributions of the extra scalars are
not relevant at one loop. In order to appreciate this effect 
notice the differences between Fig.~\ref{fig:parameter1}a 
and Fig.~\ref{fig:parameter1}c  where the gaugino mass changes 
from $200$\,GeV to $1$\,TeV and the unification scale 
for $M_{\Sigma_3}=M_{\Sigma_8}$ is always above $10^{17}$\,GeV 
in the latter. In all scenarios shown in Fig.~\ref{fig:parameter1} 
unification scale can be naturally at the string scale~\cite{BFY}.

\begin{figure*}
\begin{center}
\epsfig{file=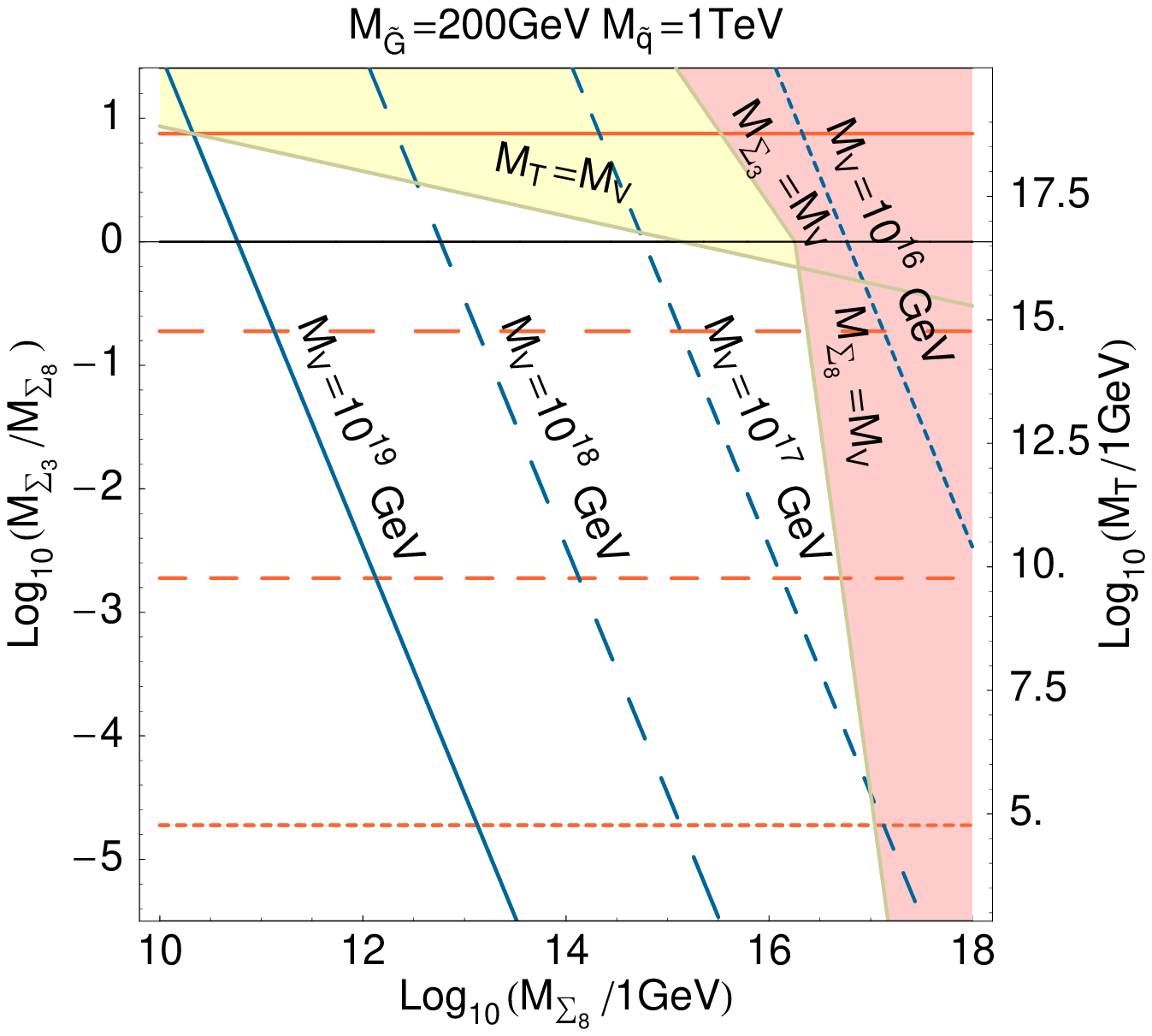,width=10cm} \caption{The whole parameter
space for gauge coupling unification for light $M_T$ scenario in
the $M_{\Sigma_3}$-$M_{\Sigma_8}$ plane. Input parameters are the
same as in Fig.~\ref{fig:parameter1}a.} \label{fig:lightmt}
\end{center}
\end{figure*}

We recall that the triplets could be light once their couplings to
matter are suppressed~\cite{Dvali}. To stress that point we show
the possibility to achieve unification in this case in 
Fig.~\ref{fig:lightmt}. It is generated with the same values of
input parameters as Fig.~\ref{fig:parameter1}a but this time we
include the region where both $T$ and $\hat{T}$ are very light.
That region could be probed at future colliders, particularly at
the LHC. For relevant signals at future colliders see 
reference~\cite{Cheung} .

Now, let us study the unification constraints at two-loop level. 
At the two-loop level, $M_T$ picks up dependence on the absolute
mass scale of $\Sigma_{8(3)}$. This means that if one considers
$M_{\Sigma_3}=M_{\Sigma_8}$ case and allows 
$M_{\Sigma_8}(=M_{\Sigma_3})$ to be below $M_T$ one can establish
an upper bound on $M_T$ \textit{only\/} after one imposes a
condition that $M_{GUT}$ is below some natural cutoff. For
example, if we take that cutoff to be $M_\mathrm{Planck}$, the
correct bound for $\tan \beta =4$, $M_{\tilde G}=1$\,TeV and
$M_{\tilde q}=1$\,TeV reads $M_T^0 < 1.2 \times 10^{16}$\,GeV.
Here, $M_T^0$ represents the mass of the triplet for the case when
$M_{\Sigma_3}=M_{\Sigma_8}$. In fact, we find
\begin{equation}
\label{limit} 5.13^{+5.07}_{-2.63} \times 10^{15}\,\textrm{GeV}
\leq M_T^0 \leq 1.2^{+1.18}_{-0.61} \times 10^{16}\,\textrm{GeV},
\end{equation}
in order for successful unification to take place. This range of
values for $M_T^0$ is obtained by allowing for arbitrary absolute
mass scale of degenerate $\Sigma_3$ and $\Sigma_8$ fields. For the
values quoted in Eq.~\eqref{limit} this scale varies between $7.3
\times 10^{10}$\,GeV for the upper bound and $1.5 \times 10^{16}$\,GeV
for the lower bound, respectively. Obviously, the constraint on
$M_T^0$ is rather tight. Quoted uncertainties in Eq.~\eqref{limit}
reflect $1\sigma$ uncertainty in $\alpha_s(M_Z)_{\overline{MS}}=0.1176 \pm 0.0020$ 
as given in~\cite{pdg}. We stress again that the upper bound 
is cutoff dependent.

\begin{figure*}
\begin{center}
\epsfig{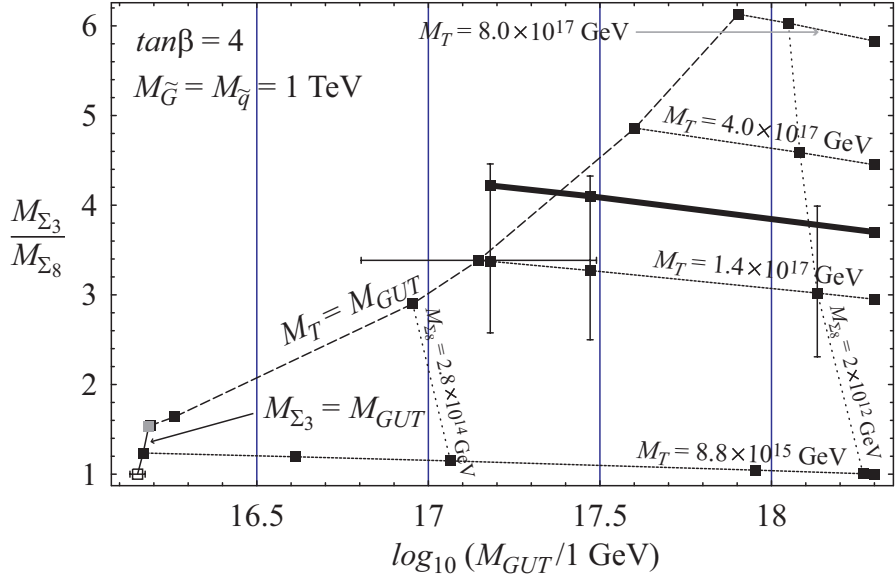} \caption{Parameter space
for successful gauge coupling unification at the two-loop level in
the $M_{GUT}$-$M_{\Sigma_3}/M_{\Sigma_8}$ plane. Lines of constant
$M_T$ and $M_{\Sigma_8}$ are shown. The dashed line corresponds to
$M_T = M_{GUT}$. Points (boxes) are exact numerical solutions for
central value of input parameters while lines represent
appropriate linear interpolation. Input parameters are specified
in detail in the text as well as the origin of shown error bars.}
\label{fig:twoloop}
\end{center}
\end{figure*}

If we depart from the $M_{\Sigma_3}=M_{\Sigma_8}$ assumption then
the upper limit on $M_T$ basically corresponds to the cutoff of 
the theory. We opt to present the two-loop analysis in
Fig.~\ref{fig:twoloop} as a contour plot of the masses of
$\Sigma_8$ and Higgs triplet fields in the 
$M_{GUT}$-$M_{\Sigma_3}/M_{\Sigma_8}$ plane. In order to generate
it we use values for gauge couplings at $M_Z$ taken from
Ref.~\cite{pdg}. In addition, we take into account CKM mixing 
parameters as given in Ref.~\cite{pdg} and include the 
effect of all three families. We consider only the $\tan \beta =4$ 
case. In our analysis the exact numerical solutions is generated 
for sufficient number of points to have smooth interpolation.

Some comments are in order with respect to Fig.~\ref{fig:twoloop}.
Vertical errors on the points that give $M_T=1.4 \times
10^{17}$\,GeV line correspond to the case when $\alpha_s
(M_Z)_{\overline{MS}}$ is varied within $1\sigma$ while $M_{GUT}$
and $M_T$ are kept fixed. In other words, what is varied there are
the ratio $M_{\Sigma_3}/M_{\Sigma_8}$ and $M_{\Sigma_8}$.
Horizontal errors on one of the points on the $M_T=1.4 \times
10^{17}$\,GeV line also correspond to the $1\sigma$ variation in
$\alpha_s (M_Z)_{\overline{MS}}$. This time $M_{\Sigma_8}$ and
$M_{GUT}$ are varied with $M_{\Sigma_3}/M_{\Sigma_8}$ and $M_T$
held fixed. Both types of error bars are given to demonstrate the
impact of the least experimentally known input parameter. The
dependence on $\tan \beta$ is practically negligible.

The only other major dependence of $M_T$ is on the exact spectrum
of the SUSY particles here encoded in parameters $M_{\tilde G}$
and $M_{\tilde q}$ for simplicity. This dependence can be treated
in a satisfactory manner only if and when this spectrum is
experimentally establish and/or better constrained. In any case,
if we assume $M_{\tilde G}=M_{\tilde q}=500$\,GeV instead of
$M_{\tilde G}=M_{\tilde q}=1$\,TeV then $M_T=1.4 \times
10^{17}$\,GeV line in Fig.~\ref{fig:twoloop} is given by the thick
line. As one can see, the ``overall'' change in the SUSY scale
within the region that can be directly probed in experiments is
still less significant than the uncertainty in $\alpha_s$.

There is however one important point that regards exact SUSY
spectrum that we need to address. Namely, it is well known that
the GUT scale unification of gaugino masses implies that
$M_1/\alpha_1=M_2/\alpha_2=M_3/\alpha_3$ at any given scale up to
small corrections. Here $M_1$, $M_2$ and $M_3$ are the 
Bino, Wino and gluino masses, respectively. We can thus 
infer that at low-scale gaugino unification predicts 
$M_3/M_2 \simeq 3.5$. This, on the other hand, is obviously in 
conflict with our assumption of gaugino degeneracy. It is a 
simple exercise to show that, at the one-loop level, 
$M_T$ scales as $(M_2/M_3)^{5/3}$. So, $M_T$ dependence 
on the ratio of relevant gaugino masses is less severe
than on the $M_{\Sigma_3}$-$M_{\Sigma_8}$ mass splitting. In other
words, we have captured the dominant effect that controls
predictions for $M_T$ within the minimal framework. But, if one
assumes that this pattern of gaugino masses is indeed the correct
one it is easy to apply it to our analysis. For example, the
limits that are quoted in Eq.~\eqref{limit} should be divided by
8. Only then our results could be compared with the existing
results in the literature such as the one in Ref.~\cite{Murayama}.

Fig.~\ref{fig:twoloop} covers the parameter space of the minimal
SUSY $SU(5)$ in both renormalizable and nonrenormalizable case and
it extends only to $M_{GUT} < 2 \times 10^{18}$\,GeV. We show the 
explicit dependence of $M_T$ on all relevant parameters, including
$M_{GUT}$ and absolute scale of $M_{\Sigma_8}$. Gray box in
Fig.~\ref{fig:twoloop} marks a point at which
$M_{\Sigma_3}=M_T=M_{GUT}$, while empty box represents a point
where $M_{\Sigma_3}=M_{\Sigma_8}=M_{GUT}$. Note that the error
bars at the latter point correspond to the $1\sigma$ variation of
$\alpha_s$ as shown on the left side of Eq.~\eqref{limit}. The
error bars are much smaller there with respect to error bars
elsewhere since only $M_T^0$ is allowed to vary to generate
successful unification.

Last point we want to discuss is the fact that more recent lattice
QCD evaluations~\cite{JLQCD,Aoki:2004xe,Aoki:2006ib} of the proton 
decay matrix element consistently imply a value that is larger
than the value ($\alpha_H=0.003$\,GeV$^3$) used to derive the limit
$M_T > 1.4 \times 10^{17}$\,GeV. Ref.~\cite{Murayama} offers
alternative, even more stringent experimental limit on $M_T$,
based on the value $\alpha_H=0.014$\,GeV$^3$~\cite{JLQCD}. When we
update it to incorporate the latest experimental bound on the partial
proton lifetime it reads $M_T > 3.7 \times 10^{17}$\,GeV. One 
can see from Fig.~\ref{fig:twoloop} that even this limit 
can be satisfied with $M_{\Sigma_3}/M_{\Sigma_8} \geq 4$.
As far as the correctness of the Eq.~\eqref{borut} is concerned,
at the two-loop level, for the central values of our set of input
parameters, it should read
\begin{equation}
M_{\Sigma_3} > 2.6 M_{\Sigma_8},
\end{equation}
in order to have $M_T > 1.4 \times 10^{17}$\,GeV.
\subsection{$d=6$ proton decay in SUSY $SU(5)$}
Let us investigate the predictions for proton decay. Since the
$d=6$ contributions are the least model dependent we focus on
them. Assuming $Y_U=Y_U^T$~\cite{d6} the decay rates due to the
presence of the superheavy gauge bosons are:
\begin{equation}
\Gamma(p \to \pi^+ \bar{\nu})= \frac{\pi \ m_p \ \alpha_{GUT}^2}
{2 \ f_{\pi}^2 \ M_{GUT}^4} \ A_L^2 \ |\alpha_H|^2 \ (1 \ + \ D \ + \ F)^2 \
|V_{CKM}^{11}|^2~,
\end{equation}
\begin{equation}
\Gamma(p \to K^+ \bar{\nu})= \frac{\pi \ \alpha_{GUT}^2}
{2 \ f_{\pi}^2 \ M_{GUT}^4} \ \frac{(m_p^2 - m_K^2)^2}{m_p^3}
\ A_L^2 \ |\alpha_H|^2 \
\left[ A_1^2 \ |V_{CKM}^{11}|^2 \ + \ A_2^2 \ |V_{CKM}^{12}|^2\right]^2~,
\end{equation}
where
\begin{eqnarray}
A_1&=& \frac{2}{3} \frac{m_p}{m_B} D~, \nonumber\\
A_2 &=& 1 \ + \ \frac{1}{3} \frac{m_p}{m_B} (D + 3 F)~.
\end{eqnarray}
In the above equations $m_B$ is the average baryon mass.
$D$, $F$ and $\alpha_H$ are the parameters of the chiral Lagrangian,
while $A_L$ takes into account the renormalization effects.
See reference~\cite{review} for details.

Let us study the impact of the unification constraints studied before 
on the proton lifetime. We have noticed that in all scenarios 
showed in Fig.~\ref{fig:parameter1} the unification scale 
is always $M_{GUT} \gtrsim 10^{16}$ GeV, therefore we can set a 
lower bound on the partial proton lifetimes:
\begin{eqnarray}
\tau(p \to \pi^+ \bar{\nu}) & \gtrsim &  \ 8 \times 10^{35}\,\textrm{years}~, \\
\tau(p \to K^+ \bar{\nu}) & \gtrsim & \ 7.6 \times
10^{37}\,\textrm{years}~.
\end{eqnarray}
Here we have used $\alpha_H=0.015 \ \textrm{GeV}^3$~\cite{Aoki:2004xe,Aoki:2006ib}. 
Notice that those lower bounds are very conservative, and valid for a minimal
realistic supersymmetric $SU(5)$ with $Y_U=Y_U^T$. The values above tell 
us that if we want to test the predictions coming from $d=6$ operators in 
minimal supersymmetric $SU(5)$ in the next generation of proton decay 
experiments the lower bounds have to be improved by at least four orders 
of magnitude. For new proposals of proton decay experiments 
see Ref.~\cite{experiments}.

Now, let us discuss the correlation between the unification 
constraints, and the predictions for nucleon decay coming 
from the $d=5$ and $d=6$ contributions. We have argued that in the realistic SUSY 
$SU(5)$ there is no a well-defined lower bound on $M_T$. However, if 
we stick to the strongest bound on $M_T$, $M_T > 10^{17}$ GeV, 
coming from $d=5$ proton decay we can conclude that in this case 
the mass of the superheavy gauge bosons is always, $M_V > 10^{17}$ GeV 
and the partial proton decay lifetimes read 
$\tau(p \to \pi^+ \bar{\nu})  \gtrsim   \ 8 \times 10^{39}\,\textrm{years}$, 
and $\tau(p \to K^+ \bar{\nu})  \gtrsim  \ 7.6 \times 10^{41}\,\textrm{years}$. 
In this case there is no hope to test the minimal supersymmetric 
$SU(5)$ at future proton decay experiments.

\section{Minimal SUSY Flipped $SU(5)$: Unification versus nucleon decay}
There exists another possibility to use $SU(5)$ to partially unify
the SM matter fields. In that approach the electric charge which
is a generator of the conventional $SU(5)$ is taken to be a linear
combination of generators operating in both $SU(5)$ and an extra
$U(1)$. This approach leads to a so-called flipped
$SU(5)$~\cite{DeRujula,Barr,Derendinger,Antoniadis}. The matter
unifies but its embedding in $SU(5)$ differs with respect to the
ordinary $SU(5)$ assignment; it can be obtained by the following
flip: $d^C \leftrightarrow u^C$, $e^C \leftrightarrow \nu^C$, $u
\leftrightarrow d$ and $\nu \leftrightarrow e$. Since the matter
unification differs from what one has in ordinary $SU(5)$, the
proton decay predictions are also different~\cite{Barr}. In what
follows we will investigate unification constraints on the mass
spectrum of the minimal flipped $SU(5)$ and corresponding
implications for proton decay signatures. For recent studies in
this context see reference~\cite{Murayama,Ellis}. 
However, in those previous studies the authors did not studied 
in detail the possibility to test this model at future proton 
decay experiments. 
  
Unlike ordinary $SU(5)$ the minimal flipped $SU(5)$ does not
require adjoint Higgs but ${\bf 10}$ and $\overline{{\bf 10}}$ of
Higgs to break down to the SM. These two representations
economically implement the so-called ``missing partner mechanism''
that efficiently suppresses $d=5$ proton decay
operators~\cite{Antoniadis}. Due to that the dominant contribution
to proton decay amplitudes comes from the gauge boson exchange.
Hence, the only relevant scale for proton decay is set by their
mass $M_{V'}$.

In order to constrain $M_{V'}$ we use gauge coupling unification.
Since only $SU(2)$ and $SU(3)$ are fully embedded in $SU(5)$ we
accordingly require unification of $\alpha_2$ and $\alpha_3$ only.
The relevant equations are 
\bea 
\label{first} 
\alpha_2^{-1} (M_Z)
&=& \alpha_\gut^{-1} + \frac{1}{2\pi} \left( - \frac{19}{6}  \ln
\frac{M_\gut}{M_Z} +  \frac{13}{6}  \ln \frac{M_\gut}{M_{\tilde
q}} +  2             \ln \frac{M_\gut}{M_{\tilde G}} -  6
\ln \frac{M_\gut}{M_{V'}} \right)~, \nn
\\
\alpha_3^{-1} (M_Z) &=& \alpha_\gut^{-1} + \nn \\ &+&
\frac{1}{2\pi} \left( - 7             \ln \frac{M_\gut}{M_Z} +  2
\ln \frac{M_\gut}{M_{\tilde q}} +  2             \ln
\frac{M_\gut}{M_{\tilde G}} -  4             \ln
\frac{M_\gut}{M_{V'}}
+           2     \ln{\frac{M_\gut}{M_{T'}}} \right)~, \nn \\
\label{flippedsusy} 
\eea 
where $M_{T'}$ is a common mass for the triplet Higgs fields. 
The number of triplets in flipped $SU(5)$ is twice the number 
of triplets in ordinary $SU(5)$. That is the reason why 
our Eqs.~\eqref{first} defer from what has been presented 
in Ref.~\cite{Murayama}. In any case, the solution to these 
equations is 
\beq \label{triplet} M_{T'} = \frac{M_Z^{23/12}
\, M_{\tilde q}^{1/12}}{ M_{V'}} \, \exp \left[ \pi \left(
\alpha_2^{-1} (M_Z)- \alpha_3^{-1} (M_Z) \right) \right]~. 
\eeq
Again, we consider the case $M_{V'} = M_{GUT}$, where $M_{GUT}$
represents the scale where $\alpha_2$ and $\alpha_3$ unify. Note
that Eq.~\eqref{triplet} is valid as long as $M_{T'} \leq
M_{GUT}$. This immediately yields 
\beq 
\label{bound} 
M_{V'} \equiv
M_{GUT} \geq M_Z \exp \left[ \frac{\pi}{2} \left(
\alpha_2^{-1}(M_Z) - \alpha_3^{-1}(M_Z) \right) \right], 
\eeq
where we have used $M_{\tilde q} \ge M_Z$. Notice that the
dependence on $M_{\tilde q}$ is extremely weak. The right-hand
side of Eq.~\eqref{bound} contains only the low-energy input which
is well-known. Upon inserting the latest experimental values we
obtain the lower bound $M_{V'} \geq 2.34 \times 10^{16}$\,GeV.
This bound allows us to set a lower bound on proton lifetime as we
show later.

\begin{figure*}
\begin{center}
\epsfig{file=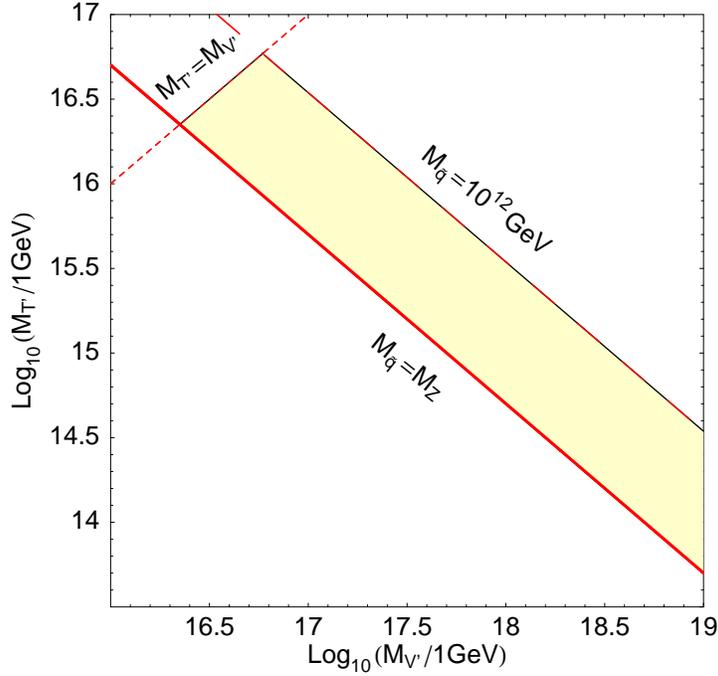,width=9.5cm} \caption{Viable parameter
space (shaded region) of the minimal flipped SUSY $SU(5)$ in the
$M_{V'}$--$M_{T'}$ plane in the $\overline{DR}$ scheme. Same input
values as in Fig.~\ref{fig:parameter1}.} \label{fig:flipped}
\end{center}
\end{figure*}
The available parameter space in the minimal flipped $SU(5)$ for
the exact unification of $\alpha_2$ and $\alpha_3$ in the
$M_{V'}$--$M_{T'}$ plane is shown in Fig.~\ref{fig:flipped}. The
allowed region is bounded by the constraints $M_Z \le
M_{\tilde{q}} \le 10^{12}$~GeV, and $M_{T'} \le M_{V'}$. The upper
bound on $M_{\tilde{q}}$ is coming from the cosmological limit on
the gluino lifetime~\cite{cosmo}. Notice that as in the case of
minimal SUSY $SU(5)$ the unification of $\alpha_2$ and $\alpha_3$
could be at the string scale.

Finally, let us incorporate the GUT scale unification of gaugino
masses in the minimal flipped $SU(5)$ context. This time it
implies $M_2/\alpha_2=M_3/\alpha_3$ at any given scale up to small
corrections. Simple exercise yields 
\beq \label{bound1} M_{GUT}
\geq M_Z^{23/24} M_{\tilde q}^{1/24} \left(\frac{M_{\tilde
H}}{M_3}\right)^{1/6} \left(\frac{M_2}{M_3}\right)^{1/3} \exp
\left[ \frac{\pi}{2} \left( \alpha_2^{-1}(M_Z) -
\alpha_3^{-1}(M_Z) \right) \right], 
\eeq 
where $M_{\tilde H}$ stands for the Higgsino masses. 
The dependence on the ratio of relevant gaugino masses 
is much weaker than in the ordinary $SU(5)$. And, unlike 
in ordinary $SU(5)$, $M_T$ does depend on the absolute scales 
of $M_2$ and $M_3$. The heavier $M_3$ is the lighter $M_{GUT}$ 
becomes although this dependence is also very weak. It is thus 
easy to see that possible gaugino unification has basically 
no impact on the $M_{V'}$ bound we quoted earlier.
\subsection{$d=6$ proton decay in flipped SUSY $SU(5)$}
Let us now turn our attention to proton decay signatures. In
Ref.~\cite{distinguishing} it has been shown the possibility to
make clean tests of minimal flippled $SU(5)$ through the channel
\beq 
\label{xxx5} 
\Gamma(p \rightarrow \pi^+ \bar{\nu})= k^4 \
C_2~, 
\eeq 
and the ratio between different channels with charged
antileptons 
\beq 
\label{flippedK0} 
\frac{\Gamma (p \to K^0 e^+)}
{\Gamma (p \to \pi^0 e^+)} = \frac{\Gamma (p \to K^0 \mu^+)}
{\Gamma (p \to \pi^0 \mu^+)}
        = 2 \frac{C_3}{C_2} \ \frac{\left|V_{CKM}^{12}\right|^2}
{\left|V_{CKM}^{11}\right|^2}=0.018, \eeq where $k= g_5/\sqrt{2}
{M^{-1}_{V'}}$, and \bea C_2 &=& m_p A_L^2 \left|\alpha_H\right|^2
(1+D+F)^2 /
8\pi f_{\pi}^2~, \nonumber \\
C_3&=& \frac{(m_p^2-m_K^2)^2}{8 \pi f_\pi^2 m_p^3}  A_L^2
        \left|\alpha_H\right|^2 \left[1+{\frac{m_p}{m_B}} (D-F)\right]^2~.
\eea
Recall that the Yukawa matrix for down quarks is symmetric in
minimal flipped $SU(5)$, and $\Gamma(p \to K^+\bar{\nu}) =
\Gamma(n \to K^0 \bar{\nu}) = 0$. See reference~\cite{review} for
details and the values of the different constants in the equations
above.

To set a bound on the partial proton lifetime we need the value 
of $\alpha_{GUT}(=g_5^2/(4 \pi))$ that corresponds to the minimal 
GUT scale. This can be obtained from
Eq.~\eqref{first}: 
\beq
\alpha_\gut^{-1}=\alpha_2^{-1}(M_Z)-\frac{1}{2 \pi} \ln
\frac{M_\gut}{M_Z}.
\eeq 
Taking into account our lower bound on $M_\gut$ we 
obtain $\alpha_\gut^{-1} \leq 24.24$. Now, using
$\alpha_H=0.015 \ \textrm{GeV}^3$ the lower bound on the partial
proton lifetime is 
\beq \label{lower} \tau(p \rightarrow \pi^+
\bar{\nu})\geq \ 2.15 \times 10^{37} \ \textrm{years}~. 
\eeq 
Using this lower bound we can obtain a lower bound on the partial proton
lifetime for the channels with charged antileptons~\cite{distinguishing}: 
\beq 
\tau(p \to \pi^0 e^+
(\mu^+)) > 2 \, \tau(p \to \pi^+ \bar{\nu})~. 
\eeq 
As in the previous section, we conclude that it will be very difficult to
test the minimal flipped $SU(5)$ at the next generation of proton
decay experiments by looking at the channel $p \to \pi^+
\bar{\nu}$ and using the ratio given in Eq.~(\ref{flippedK0}). 
We recall that in the near future the lower bounds will be improved
by two or three orders of magnitude. If proton decay is found, in
particular the channels above, and the lifetime is below this
lower bound then minimal flipped SUSY $SU(5)$ will be highly
disfavored.

\section{Conclusions}
We have investigated the unification constraints in the minimal 
supersymmetric grand unified theories based on $SU(5)$. 
The most general constraints on the spectrum of minimal 
supersymmetric $SU(5)$ and flipped $SU(5)$ have been shown. 
The upper bound on the mass of the colored Higgs mediating 
proton decay has been discussed in detail. We studied the issue of 
proton decay in both GUT models, pointing out lower bounds 
on the partial proton lifetime for the relevant channels 
in flipped SUSY $SU(5)$. We conclude that if proton decay 
is found in the next generation of experiments with a 
lifetime lower than $10^{37}$ years, then flipped $SU(5)$ 
could be excluded as a realistic GUT candidate. 
In the case of the minimal supersymmetric $SU(5)$ we have shown 
that if we stick to the strongest bound on the colored triplet mass 
coming from dimension five proton decay contributions there 
is no hope to test the model at future nucleon decay experiments 
through the dimension six operators.    
\section*{Acknowledgements}
{\small The work of G.~R. was partially supported by Ministerio de
Educaci\'on y Ciencia (MEC) under grant FPA2004-00996,
Generalitat Valenciana under grant GV05-015,
Consejo Superior de Investigaciones Cient\'{\i}ficas (CSIC)
under grant PIE 200650I247, and European Commission FLAVIAnet
MRTN-CT-2006-035482. P.~F.~P. has been
supported by {\em Funda\c{c}\~{a}o para a Ci\^{e}ncia e a
Tecnologia} (FCT, Portugal) through the project CFTP,POCTI-SFA-2-777
and a fellowship under project POCTI/FNU/44409/2002. P.~F.~P would
like to thank the Instituto de F\'{\i}sica Corpuscular (IFIC) in
Valencia for hospitality and B. Bajc and G. Senjanovi\'c for discussions.}



\begin{thebibliography}{99}

\bibitem{GG}
  H.~Georgi and S.~L.~Glashow,
  ``Unity Of All Elementary Particle Forces,''
  Phys.\ Rev.\ Lett.\  {\bf 32} (1974) 438.

\bibitem{Pati-Salam}
  J.~C.~Pati and A.~Salam,
  ``Is Baryon Number Conserved?,''
  Phys.\ Rev.\ Lett.\  {\bf 31} (1973) 661.

\bibitem{Langacker}
  P.~Langacker,
  ``Grand Unified Theories And Proton Decay,''
  Phys.\ Rept.\  {\bf 72} (1981) 185.

\bibitem{review}
P.~Nath and P.~Fileviez~P\'erez,
``Proton stability in grand unified theories, in strings, and in
branes,''
  arXiv:hep-ph/0601023.

\bibitem{non-SUSY-SU(5)}
  I.~Dorsner and P.~Fileviez~P\'erez,
   ``Unification without supersymmetry: Neutrino mass, proton decay and  light leptoquarks,''
  Nucl.\ Phys.\ B {\bf 723} (2005) 53
  [arXiv:hep-ph/0504276];
  I.~Dorsner, P.~Fileviez~P\'erez and R.~Gonzalez Felipe,
  ``Phenomenological and cosmological aspects of a minimal GUT scenario,''
  Nucl.\ Phys.\ B {\bf 747} (2006) 312
  [arXiv:hep-ph/0512068].

\bibitem{non-SUSY-SU(5)I}
  I.~Dorsner, P.~Fileviez~P\'erez and G.~Rodrigo,
  ``Fermion masses and the UV cutoff of the minimal realistic SU(5),''
  arXiv:hep-ph/0607208.

\bibitem{rotating}
  I.~Dorsner and P.~Fileviez P\'erez,
  ``Could we rotate proton decay away?,''
  Phys.\ Lett.\ B {\bf 606} (2005) 367
  [arXiv:hep-ph/0409190].

\bibitem{upper}
  I.~Dorsner and P.~Fileviez P\'erez,
  ``How long could we live?,''
  Phys.\ Lett.\ B {\bf 625} (2005) 88
  [arXiv:hep-ph/0410198].

\bibitem{SUSYSU(5)}
  S.~Dimopoulos and H.~Georgi,
  ``Softly Broken Supersymmetry And SU(5),''
  Nucl.\ Phys.\ B {\bf 193} (1981) 150;
N.~Sakai,
  ``Naturalness In Supersymmetric 'Guts',''
  Z.\ Phys.\ C {\bf 11} (1981) 153.


\bibitem{Murayama}
  H.~Murayama and A.~Pierce,
  ``Not even decoupling can save minimal supersymmetric SU(5),''
  Phys.\ Rev.\ D {\bf 65} (2002) 055009
  [arXiv:hep-ph/0108104].


\bibitem{Nath-pd1}
  P.~Nath, A.~H.~Chamseddine and R.~Arnowitt,
  ``Nucleon Decay In Supergravity Unified Theories,''
  Phys.\ Rev.\  D {\bf 32} (1985) 2348.

\bibitem{Nath-pd2}
  R.~Arnowitt, A.~H.~Chamseddine and P.~Nath,
  ``Nucleon Decay Branching Ratios In Supergravity SU(5) Guts,''
  Phys.\ Lett.\  B {\bf 156} (1985) 215.

\bibitem{Hisano}
  J.~Hisano, H.~Murayama and T.~Yanagida,
  ``Nucleon decay in the minimal supersymmetric SU(5) grand unification,''
  Nucl.\ Phys.\ B {\bf 402} (1993) 46
  [arXiv:hep-ph/9207279].

\bibitem{Ellis:1979fg}
  J.~R.~Ellis and M.~K.~Gaillard,
  ``Fermion Masses And Higgs Representations In SU(5),''
  Phys.\ Lett.\ B {\bf 88} (1979) 315.


\bibitem{Dvali}
  G.~R.~Dvali,
   ``Can 'doublet - triplet splitting' problem be solved without doublet -
  triplet splitting?,''
  Phys.\ Lett.\ B {\bf 287} (1992) 101.

\bibitem{Bajc1}
  B.~Bajc, P.~Fileviez P\'erez and G.~Senjanovi\'c,
  ``Proton decay in minimal supersymmetric SU(5),''
  Phys.\ Rev.\ D {\bf 66} (2002) 075005
  [arXiv:hep-ph/0204311].

\bibitem{Bajc2}
  B.~Bajc, P.~Fileviez P\'erez and G.~Senjanovi\'c,
  ``Minimal supersymmetric SU(5) theory and proton decay: Where do we stand?,''
  arXiv:hep-ph/0210374; P.~Fileviez P\'erez,
``Phenomenological aspects of supersymmetric gauge theories,''
  arXiv:hep-ph/0310199.

\bibitem{Nath1}
  P.~Nath,
  ``Hierarchies and Textures in Supergravity Unification,''
  Phys.\ Rev.\ Lett.\  {\bf 76} (1996) 2218
  [arXiv:hep-ph/9512415].

\bibitem{Nath2}
  P.~Nath,
   ``Textured Minimal and Extended Supergravity Unification and Implications for
  Proton Stability,''
  Phys.\ Lett.\ B {\bf 381} (1996) 147
  [arXiv:hep-ph/9602337].

\bibitem{Berezhiani}
  Z.~Berezhiani, Z.~Tavartkiladze and M.~Vysotsky,
  ``d = 5 operators in SUSY GUT: Fermion masses versus proton decay,''
  arXiv:hep-ph/9809301.

\bibitem{David}
  D.~Emmanuel-Costa and S.~Wiesenfeldt,
  ``Proton decay in a consistent supersymmetric SU(5) GUT model,''
  Nucl.\ Phys.\ B {\bf 661} (2003) 62
  [arXiv:hep-ph/0302272].

\bibitem{Shafi}
  C.~T.~Hill,
   ``Are There Significant Gravitational Corrections To The Unification
  Scale?,''
  Phys.\ Lett.\ B {\bf 135} (1984) 47. Q.~Shafi and C.~Wetterich,
   ``Modification Of GUT Predictions In The Presence Of Spontaneous
  Compactification,''
  Phys.\ Rev.\ Lett.\  {\bf 52} (1984) 875.

\bibitem{pdg}
  W.~M.~Yao {\it et al.}  [Particle Data Group],
  ``Review of particle physics,''
  J.\ Phys.\ G {\bf 33} (2006) 1.

\bibitem{Barbieri}
  R.~Barbieri and L.~J.~Hall,
  ``Grand unification and the supersymmetric threshold,''
  Phys.\ Rev.\ Lett.\  {\bf 68} (1992) 752;
J.~Hisano, H.~Murayama and T.~Yanagida,
  ``Probing GUT scale mass spectrum through precision measurements on the weak
  scale parameters,''
  Phys.\ Rev.\ Lett.\  {\bf 69} (1992) 1014.


\bibitem{splitsusy}
  N.~Arkani-Hamed and S.~Dimopoulos,
   ``Supersymmetric unification without low energy supersymmetry and  signaturesfor fine-tuning at the LHC,''
  JHEP {\bf 0506} (2005) 073
  [arXiv:hep-th/0405159]; 
G.~F.~Giudice and A.~Romanino,
  ``Split supersymmetry,''
  Nucl.\ Phys.\ B {\bf 699} (2004) 65
  [Erratum-ibid.\ B {\bf 706} (2005) 65]
  [arXiv:hep-ph/0406088].

\bibitem{BFY}
  C.~Bachas, C.~Fabre and T.~Yanagida,
  ``Natural gauge-coupling unification at the string scale,''
  Phys.\ Lett.\ B {\bf 370} (1996) 49
  [arXiv:hep-th/9510094].

\bibitem{Cheung}
  K.~Cheung and G.~C.~Cho,
  ``TeV colored Higgsinos in alternative grand unified theories,''
  Phys.\ Rev.\ D {\bf 69} (2004) 017702
  [arXiv:hep-ph/0306068].


\bibitem{JLQCD}
  Y.~Kuramashi  [JLQCD Collaboration],
  ``Nucleon decay matrix elements from lattice QCD,''
  arXiv:hep-ph/0103264.

\bibitem{Aoki:2004xe}
  Y.~Aoki  [RBC Collaboration],
  ``Nucleon decay matrix elements with N(f) = 0 and 2 domain-wall quarks,''
  Nucl.\ Phys.\ Proc.\ Suppl.\  {\bf 140} (2005) 405
  [arXiv:hep-lat/0409114].

\bibitem{Aoki:2006ib}
  Y.~Aoki, C.~Dawson, J.~Noaki and A.~Soni,
  ``Proton decay matrix elements with domain-wall fermions,''
  arXiv:hep-lat/0607002.

\bibitem{d6}
  P.~Fileviez P\'erez,
  ``Fermion mixings vs d = 6 proton decay,''
  Phys.\ Lett.\ B {\bf 595} (2004) 476
  [arXiv:hep-ph/0403286].

\bibitem{experiments}
  Talks given by S.~Katsanevas , K.~Nakamura, R.~Wilson and A.~de~Bellefon
  at Workshop on Next generation Nucleon decay and Neutrino detectors
  2006 (NNN06), http://neutrino.phys.washington.edu/nnn06/

\bibitem{DeRujula}
  A.~De Rujula, H.~Georgi and S.~L.~Glashow,
  ``Flavor Goniometry By Proton Decay,''
  Phys.\ Rev.\ Lett.\  {\bf 45} (1980) 413.

\bibitem{Barr}
  S.~M.~Barr,
  ``A New Symmetry Breaking Pattern For SO(10) And Proton Decay,''
  Phys.\ Lett.\ B {\bf 112} (1982) 219.

\bibitem{Derendinger}
  J.~P.~Derendinger, J.~E.~Kim and D.~V.~Nanopoulos,
  ``Anti - SU(5),''
  Phys.\ Lett.\ B {\bf 139} (1984) 170.

\bibitem{Antoniadis}
  I.~Antoniadis, J.~R.~Ellis, J.~S.~Hagelin and D.~V.~Nanopoulos,
  ``Supersymmetric flipped SU(5) revitalized,''
  Phys.\ Lett.\ B {\bf 194} (1987) 231.

\bibitem{Ellis}
  J.~R.~Ellis, D.~V.~Nanopoulos and J.~Walker,
  ``Flipping SU(5) out of trouble,''
  Phys.\ Lett.\ B {\bf 550} (2002) 99
  [arXiv:hep-ph/0205336].

\bibitem{cosmo}
  A.~Arvanitaki, C.~Davis, P.~W.~Graham, A.~Pierce and J.~G.~Wacker,
  ``Limits on split supersymmetry from gluino cosmology,''
  Phys.\ Rev.\ D {\bf 72} (2005) 075011
  [arXiv:hep-ph/0504210].

\bibitem{distinguishing}
  I.~Dorsner and P.~Fileviez P\'erez,
  ``Distinguishing between SU(5) and flipped SU(5),''
  Phys.\ Lett.\ B {\bf 605} (2005) 391
  [arXiv:hep-ph/0409095].

\end{thebibliography}
\end{document}